\documentclass[twocolumn,aps,pre,amsmath,amssymb,longbibliography]{revtex4-1}
\usepackage{graphicx}
\usepackage{dcolumn}
\usepackage{bm}
\usepackage{multirow}
\usepackage{amssymb}
\usepackage{amsmath}
\usepackage{graphicx}
\usepackage{xcolor}
\usepackage[colorlinks,citecolor=blue,linkcolor=red,urlcolor=blue]{hyperref}
\usepackage{indentfirst}
\usepackage{amsmath}
\usepackage{cases}
\usepackage{lineno}

\begin{document}
\title{Equilibrium and nonequlibrium scaling behaviors of localization transition in a non-Hermitian Aubry-Andr\'{e} model with onsite gain and loss}
\author{Wen-Jing Yu\textsuperscript{1,2}}
\altaffiliation{These authors contribute equally to this work.}
\author{Yue-Mei Sun\textsuperscript{1,2}}
\altaffiliation{These authors contribute equally to this work.}
\author{Xin-Yu Wang\textsuperscript{1,2}}
\author{Liang-Jun Zhai\textsuperscript{1,2,3}}\email{zhailiangjun@jsut.edu.cn}
\affiliation{\textsuperscript{1}The School of Mathematics and Physics, Jiangsu University of Technology, Changzhou 213001, China}
\affiliation{\textsuperscript{2}The Jiangsu Key Laboratory of Clean Energy Storage and Conversion, Jiangsu University of Technology, Changzhou 213001, China}
\affiliation{\textsuperscript{3}Jiangsu Physical Science Research Center, Nanjing 210093, China}
\date{\today}
\begin{abstract}
The interplay between non-Hermiticity and localization has attracted considerable interest, yet the driven dynamics of localization transitions in non-Hermitian systems with on-site gain and loss remains largely unexplored.
Here we investigate the critical scaling behavior and driven dynamics of the non-Hermitian Aubry-Andr\'e (AA) model with on-site gain and loss.
Through finite-size scaling analyses of the localization length, the inverse participation ratio (IPR), and the energy gap, we extract the critical exponents $\nu = 1.00(2)$, $s = 0.7965(2)$, and $z = 1.999(2)$.
These exponents are different from those of both the Hermitian AA model and the nonreciprocal hopping AA model, particularly the IPR exponent $s$, demonstrating that the gain-loss mechanism belongs to a distinct universality class.
For the driven dynamics, we focus on the case where the system is initially prepared in a gapless extended state and linearly driven across the critical point.
We verify that the finite-time scaling (FTS) framework remains applicable provided that the criterion $z' < r$ is satisfied, where $z' = 1.999(2)$ characterizes the gap closure in the extended phase and $r = z + 1/\nu \approx 2.999$.
The predicted FTS scaling forms for the IPR are numerically validated across a wide range of system sizes and driving rates, demonstrating that the unified scaling description can be successfully generalized to the gain-loss type non-Hermitian AA model.
Our work not only establishes the gain-loss AA model as a new universality class of localization transitions but also extends the applicability of the FTS framework to non-Hermitian systems with gapless initial states.
\end{abstract}
\maketitle

\section{Introduction}
The phenomenon of localization has attracted widespread attention since Anderson's seminal work, and has been extensively verified in diverse platforms from cold atoms to photonic lattices~\cite{Anderson1958,Abrahams1979,Lee1985,Kramer1993,Schwartz2007,Evers2008,Segev2013,Wiersma2013,Alexey2023, Thouless1974,Ekuma2015,Nguyen2022,Qi2026}.
However, localization is not exclusive to disordered systems.
In recent years, quasiperiodic systems have attracted considerable interest as fertile ground for exploring localization physics~\cite{Aubry1980,Biddle070601,HepengYao2019,Schirmann2024,Ribeiro2024,Jan2023,Cheng2023,Agrawalprl2020,Goblot2020,Roy2022}.
Among them, the Aubry-Andr\'{e} (AA) model stands out as the most paradigmatic example, exhibiting a sharp quantum phase transition from extended to localized eigenstates at a finite critical strength of the quasiperiodic potential, protected by an exact self-duality symmetry~\cite{Aubry1980,Hetenyi2025,DasSarma1988,Biddle2009,Biddle2011,Ganeshan2015,Liu2015,Wang2020,Zhou2023,Wangyn2026,Bu2022}.
In contrast to one-dimensional disordered systems, where any infinitesimal disorder leads to localization, the AA model's sharp transition at a finite critical point makes it an ideal platform for investigating localization physics, as verified in numerous experiments~\cite{DuL2025,Roati2008,Billy2008,Lahini2009,Schreiber2015,Luschen2018,Kohlert2019,An2021,Rispoli2019,Singh2015,Xiao2021,Lin2022}.

In parallel, the interplay between non-Hermiticity and localization has drawn growing interest, fueled by rapid advances in open quantum systems~\cite{Hatano1996,Hatano1998,Jiang2019,Zhai2021,Sun2024,Longhi2019a,Longhi2019b,Zeng2017,Xu2021,Xing2025,Cai2021,
Wang2025,Chen2025,Zhai2020,Zhang2025,Luo2021,Huang2020,Guo2021,Longhi2025,LiB2025,Ghatak2024,RoyS2022}.
Non-Hermitian Hamiltonians host a variety of exotic phenomena absent in Hermitian settings, such as the non-Hermitian skin effect, exceptional points, and unconventional topological phases~\cite{Ashida2020,Bergholtz2021,El-Ganainy2018,Yao2018,Gong2018,Kunst2018,Ashida2020,Bergholtz2021,El-Ganainy2018,Longhi2017,Shen2018,Kawabata2019,Xue2026,
Zhang2026,Gohsrich2025}.
In the context of localization, non-Hermiticity is typically introduced via two distinct mechanisms: nonreciprocal hopping or on-site gain and loss potentials~\cite{Jiang2019,Zhai2021,Longhi2019a,Longhi2019b}.
While both routes give rise to localization transitions accompanied by topological and real-to-complex spectral changes, their underlying behaviors are fundamentally different.
In the nonreciprocal AA model, the localized phase features a real spectrum and trivial topology, whereas the extended phase exhibits a complex spectrum and nontrivial topology~\cite{Jiang2019,Zhai2021}.
Strikingly, for the on-site gain and loss AA model, this correspondence is exactly reversed: the extended phase has a real spectrum and trivial topology, while the localized phase hosts a complex spectrum and nontrivial topology~\cite{Longhi2019a,Longhi2019b}.
Such a marked distinction naturally calls for a systematic comparative study of how these two non-Hermitian mechanisms influence localization transitions and critical phenomena.

On another front, driven dynamics across localization phase transitions has also become an active area of investigation~\cite{Dziarmaga2010,Polkovnikov2011,Gong2010,Huang2014,WangX2024,Gutierrez2023,Reichhardt2022,
SotoGarcia2026,WangX2024,Chang2026,Yang2017,Xu2020,Morales-Molina2014,Bairey2017,Modak2021}.
For the single-particle AA model, the Kibble-Zurek mechanism (KZM) offers a general framework for characterizing the nonequilibrium evolution under external driving~\cite{Reichhardt2022,Sinha2019,Tong2021}.
Going beyond the pristine AA model, a hybrid KZM has been proposed for systems where disorder and quasiperiodic potential coexist, capturing the driven critical behavior in the overlapping critical region~\cite{Bu2023,Liang2024}.
More recently, finite-time scaling (FTS)---a generalization of KZM---has been shown to capture the driven dynamics even for gapless initial states, where the conventional adiabatic-impulse approximation breaks down.~\cite{Wang2026}.
The nonequilibrium dynamics of non-Hermitian localization transitions have also received considerable attention~\cite{Li2024,Cheng2024,Xing2025,Xu2021,Zhai2022a,Zhai2022b,Sun2025a}.
Previous studies on the driven dynamics of the non-Hermitian AA model and the non-Hermitian DAA model have established that, despite Hermitian and non-Hermitian systems belonging to distinct universality classes, the KZM framework remains effective in describing driven dynamics across non-Hermitian localization transitions~\cite{Zhai2022a,Zhai2022b,Sun2025a}.
It is worth noting, however, that these investigations are exclusively based on the nonreciprocal-hopping mechanism. The driven dynamics of the on-site gain and loss AA model, by contrast, remains largely unexplored.
Given the fundamentally different spectral and topological characteristics of the gain-loss mechanism relative to its nonreciprocal counterpart, a systematic exploration of its driven dynamics is both timely and essential.

In this work, we investigate the scaling properties and driven dynamics of the localization phase transition in the AA model with on-site gain and loss.
Through a systematic analysis of both static critical behavior and nonequilibrium dynamics under linear driving, we obtain two main results.
First, we demonstrate that the gain-loss AA model belongs to a universality class distinct from both the Hermitian AA model and its nonreciprocal variant, as reflected in the differences among their critical exponents.
Second, we show that the FTS framework remains fully applicable even when the system is initially prepared in a gapless extended state, thereby extending the reach of FTS to non-Hermitian systems with gapless initial conditions.
Taken together, our findings offer a comprehensive perspective on the interplay among non-Hermiticity, localization criticality, and nonequilibrium driving, and contribute new insights to the universality classification of non-Hermitian localization transitions.

\section{\label{modelphase}Hamiltonian and phase diagram}

We consider a one-dimensional single-particle system described by the non-Hermitian AA model with on-site gain and loss.
The Hamiltonian reads~\cite{Longhi2019a,Longhi2019b}
\begin{equation}
H = -J\sum_{j=1}^{L} (c_j^\dagger c_{j+1} + \text{H.c.}) + \sum_{j=1}^{L} V_n(j) c_j^\dagger c_j,
\label{Eq:model}
\end{equation}
where $c_j^\dagger$ ($c_j$) is the creation (annihilation) operator of a particle at site $j$, $J$ is the hopping amplitude, and $L$ is the system size.
The onsite potential $V_n(j)$ is complex and takes the form $V_n(j)=V\cos(2\pi \gamma j+\phi)$, with $V$ and  $\phi$ controlling the potential strength and phase, and $\gamma$ an irrational number.
We parametrize the phase as
\begin{equation}
\phi = \theta + i h,
\label{Eq:phi}
\end{equation}
with $\theta \in [0, 2\pi)$ and $h \in \mathbb{R}$.
This model can be viewed as a complexification of the standard AA potential.
The non-Hermiticity is thus introduced through on-site gain and loss, rather than through nonreciprocal hopping.

We set $J=1$ as the unit of energy and choose $\gamma = (\sqrt{5}-1)/2$, the inverse golden ratio.
To satisfy periodic boundary conditions (PBC), we approximate $\gamma$ by rational numbers $F_{m-1}/F_m$, where $F_m$ are Fibonacci numbers and $L=F_m$~\cite{Jiang2019,Sinha2019}.
Under this approximation, the quasiperiodic potential is periodic on the finite lattice, and $\gamma$ converges to the golden ratio in the thermodynamic limit.

The localization phase transition of this model has been rigorously established in the thermodynamic limit.
It was shown analytically that the transition occurs along the critical line~\cite{Longhi2019a,Longhi2019b}
 \begin{eqnarray}
   h &=& \log(2J/V).
   \label{Eq:criticallocation}
 \end{eqnarray}
To characterize the localization properties, we employ the inverse participation ratio (IPR) of the eigenstate with the lowest real part of the energy, defined as~\cite{Bauer1990,Fyodorov1992}
\begin{equation}
{\rm{IPR}} = \frac{\sum_{j=1}^{L} ||\Psi(j)\rangle|^4}{ \sum_{j=1}^{L} ||\Psi(j)\rangle|^2 },
\label{Eq:IPR}
\end{equation}
where $|\Psi(j)\rangle$ is the amplitude of the eigenstate with the lowest real part of the energy at site $j$.
The IPR serves as a standard probe for distinguishing different phases: for an extended state, the wave function is uniformly distributed over all sites, yielding ${\rm{IPR}} \propto L^{-1}$;
for a localized state, the wave function occupies only a finite number of sites, giving ${\rm{IPR}} \propto L^0$;
and at the critical point, the wave function exhibits multifractal behavior, with ${\rm{IPR}} \propto L^{-s/\nu}$, where $s$ and $\nu$ are critical exponents characterizing the localization transition~\cite{Wei2019,Sun2024}.

\begin{figure}[tbp]
\centering
  \includegraphics[width=3.5 in,clip]{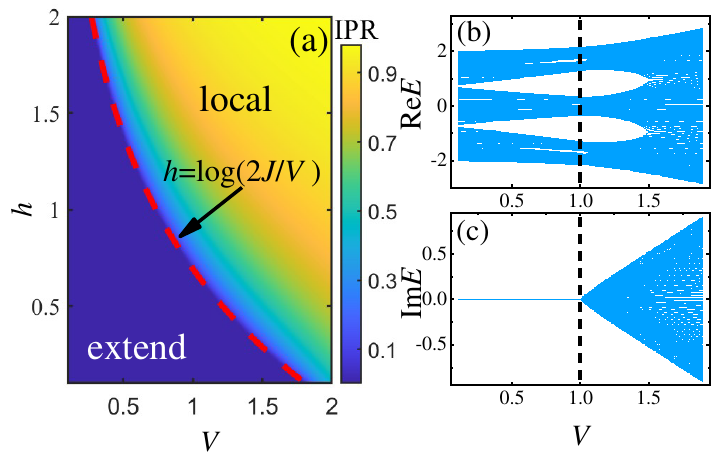}
  \vskip-3mm
  \caption{(a) Phase diagram of the gain-loss type non-Hermitian AA model in the $(V,h)$ plane, obtained from the IPR of the eigenstate with the lowest real part of the energy under PBC. The color scale represents the IPR value, where the critical line $h = \log(2J/V)$ separates the extended phase from the localized phase. (b) Real and (c) imaginary parts of the energy spectra as functions of $V$ for fixed $h = 0.693$, where the corresponding critical point is $V_c = 1$, with $L = 377$ and $\theta = 0$. Re $E$ and Im $E$ denote the real and imaginary parts. The black dashed lines in (b) and (c) mark the critical point $V_c = 1$, separating the extended phase ($V < V_c$) with purely real spectra from the localized phase ($V > V_c$) with complex spectra.}
  \label{fig:PhaseEnergy}
\end{figure}

As shown in Fig.~\ref{fig:PhaseEnergy}(a), we present the phase diagram computed from the IPR of the eigenstate with the lowest real part of the energy under PBC.
It is clearly seen that the critical line separates the localized phase from the extended phase, which confirm the analytical prediction of Eq.~\eqref{Eq:criticallocation}.

To further corroborate the connection between the localization transition and the spectral property, we present in Figs.~\ref{fig:PhaseEnergy}(b) and (c) the real and imaginary parts of the energy spectrum under PBC as functions of $V$ for a fixed $h$.
It is clearly observed that when the system is in the extended phase ($V < V_c $), where $V_c$ is the critical point of the localization transition, all eigenenergies are purely real for all states.
However, as $V$ exceeds the critical value $V_c$, imaginary parts of the eigenenergies emerge, signaling the onset of the localized phase.
This observation further confirms that, for the gain-loss AA model, the localization-delocalization transition is always accompanied by a real-to-complex spectral transition.

\begin{figure}[tbp]
\centering
  \includegraphics[width=3.3 in,clip]{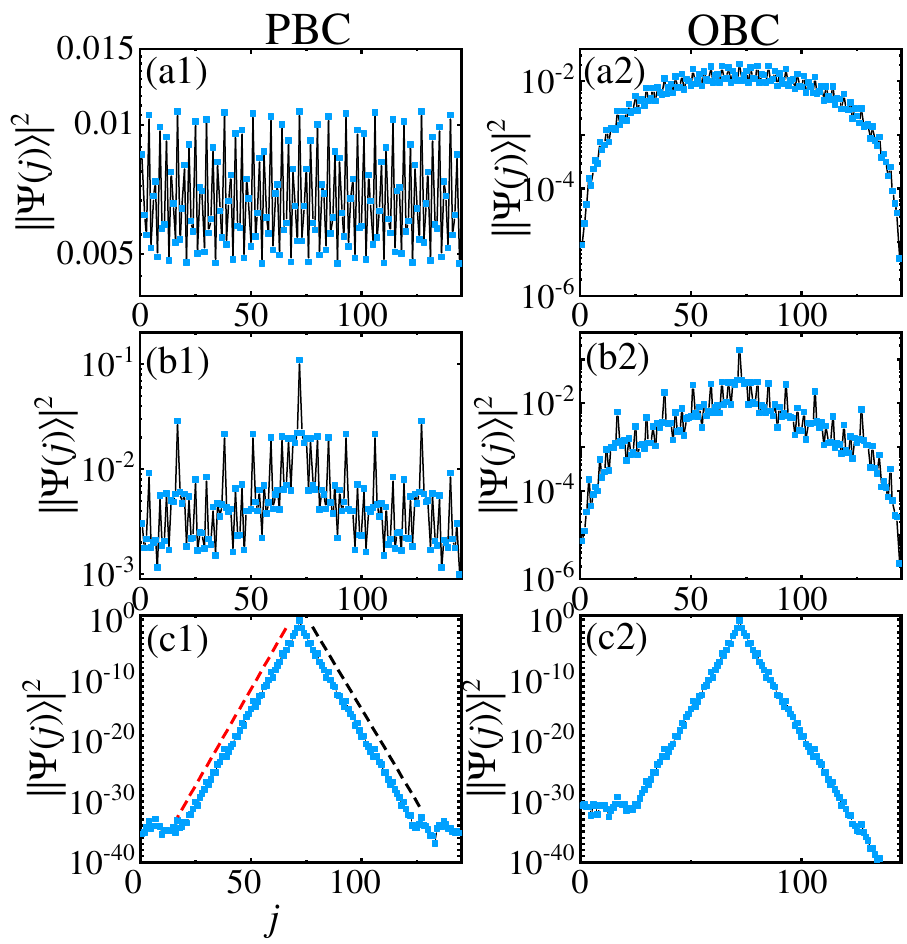}
  \vskip-3mm
  \caption{Spatial distributions of the squared wave function $|\psi(j)|^2$ for the eigenstate with the lowest real part of the energy under PBC (left column) and OBC (right column), with $L = 144$, $\theta = 0$, $h_c = 0.693$, and $V_c=1$. (a1)-(a2) Extended phase with $V = V_c - 0.5$. (b1)-(b2) Critical phase with $V = V_c$. (c1)-(c2) Localized phase with $V = V_c + 1$.
  n (c1), the black and red dashed lines are linear fits to the logarithmic wave function amplitudes on the right and left sides of the localization center, yielding slopes of $-1.4488$ and $-1.4479$, respectively, indicating nearly symmetric exponential decay.
  The wave functions are plotted on a logarithmic scale.}
  \label{fig:wavefunction}
\end{figure}

We now turn to the spatial characteristics of the eigenstates under the gain-loss mechanism. Figure~\ref{fig:wavefunction} displays the squared wave function $||\Psi(j)\rangle|^2$ under both PBC and open boundary conditions (OBC), for the extended [(a1), (a2)], critical [(b1), (b2)], and localized [(c1), (c2)] phases, respectively.
In the extended phase, as shown in Figs.~\ref{fig:wavefunction}(a1) and (a2), the wave functions under both PBC and OBC are approximately uniformly distributed across the bulk region, although the distribution near the edges is suppressed to some extent under OBC.
At the critical point, illustrated in Figs.~\ref{fig:wavefunction}(b1) and (b2), the wave functions exhibit self-similar structures under both boundary conditions, a hallmark of multifractal critical states, while the suppression near the edges becomes more pronounced under OBC.
In the localized phase, depicted in Figs.~\ref{fig:wavefunction}(c1) and (c2), the wave functions under both PBC and OBC are concentrated on a few specific sites, showing highly similar features.
Notably, we find that the decay rates on the two sides of the localization center are nearly identical.
As shown in Fig.~\ref{fig:wavefunction}(c1), a linear fit to the logarithmic wave function amplitudes yields slopes of \(-1.4488\) and \(-1.4479\) for the left and right sides, respectively, indicating symmetric exponential decay.

It is instructive to compare these features with the nonreciprocal AA model.
In that case, under OBC, the extended phase becomes a skin-effect state, where all bulk eigenstates are exponentially localized toward one boundary~\cite{Jiang2019}.
At the critical point, the wave function still preserves a self-similar structure, but with a pronounced asymmetric distribution: the probability amplitude on one side of the system is significantly enhanced compared to the other~\cite{Zhai2021}.
In the localized phase, the wave function exhibits asymmetric exponential decay, with different decay lengths on the two sides of the localization center~\cite{LiSZ2024,Tong2025}.
None of these features appear in the present gain-loss model, highlighting the fundamental difference between the two non-Hermitian mechanisms.

\section{\label{static}The static scaling behavior and critical exponent}
In this section, we study the static scaling properties of the gain-loss type non-Hermitian AA model under PBC and extract the associated critical exponents.

In addition to the IPR, we employ two other quantities to characterize the localization transition: the localization length $\xi$ and the energy gap $\Delta E$.
The localization length $\xi$ is defined as~\cite{Sinha2019}
\begin{equation}
\xi = \sqrt{\sum_{j=1}^{L} (j - j_c)^2 P_j},
\end{equation}
where $P_j = ||\Psi(j)\rangle|^2$ is the probability density at site $j$, and $j_c = \sum_j j P_j$ is the localization center.
In the localized phase, $\xi$ remains finite, whereas it diverges as $\xi \propto |\varepsilon|^{-\nu}$ when approaching the critical point, where $\varepsilon$ measures the distance to the critical point.

The energy gap $E_g$ is defined as the difference between the lowest real parts of the eigenenergies of the first excited state and the ground state:
\begin{equation}
\label{eq:Eg}
E_g = \mathrm{Re}(E_1) - \mathrm{Re}(E_0),
\end{equation}
where $E_0$ and $E_1$ are the eigenenergies of the states with the lowest and the second lowest real parts, respectively.
At the critical point, $\Delta E$ vanishes algebraically with system size as $E_g \propto L^{-z}$, providing an independent route to extract the dynamical exponent $z$.

\begin{figure}[tbp]
\centering
  \includegraphics[width=3.4 in,clip]{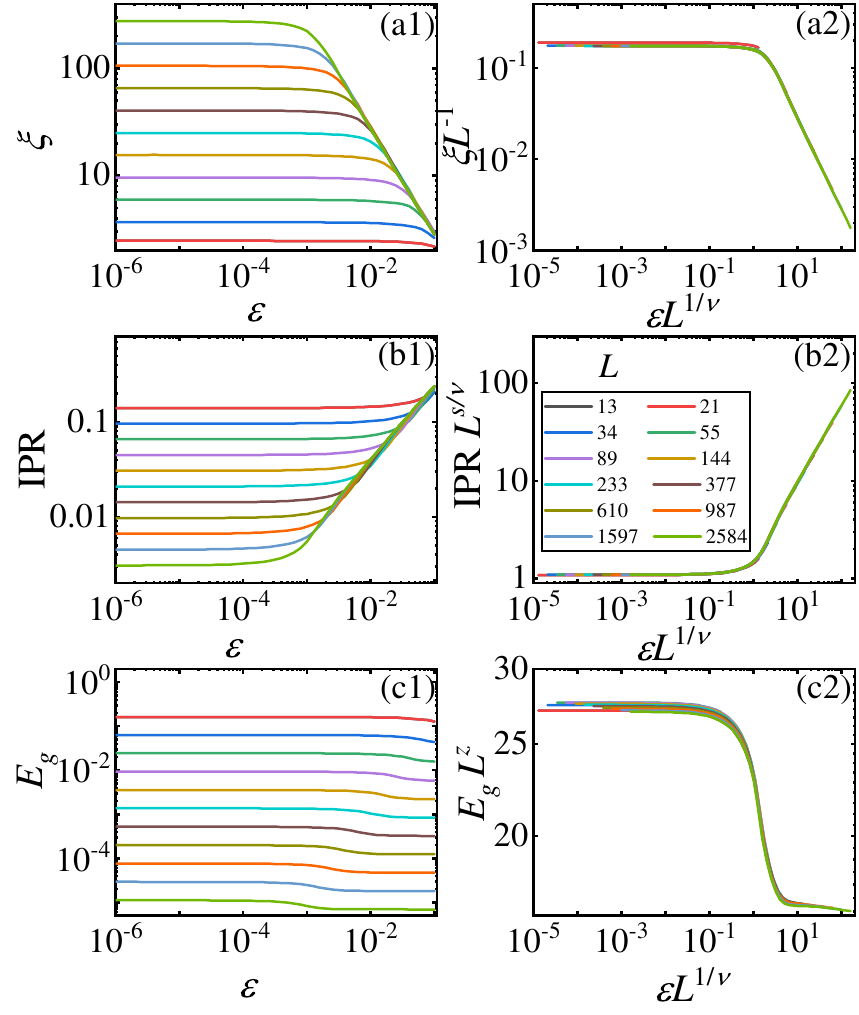}
  \vskip-3mm
  \caption{Static finite-size scaling of the three observables for the gain-loss type non-Hermitian AA model under PBC at a fixed non-Hermitian parameter $h = 0.693$. (a1) Raw data of the localization length $\xi$ as a function of $\varepsilon$ for different system sizes $L$. (a2) Rescaled curves of $\xi/L$ versus $\varepsilon L^{1/\nu}$. (b1) Raw data of the IPR as a function of $\varepsilon$ for different $L$. (b2) Rescaled curves of ${\rm IPR} \cdot L^{s/\nu}$ versus $\varepsilon L^{1/\nu}$. (c1) Raw data of the energy gap $E_g$ as a function of $\varepsilon$ for different $L$. (c2) Rescaled curves of $E_g L^z$ versus $\varepsilon L^{1/\nu}$. All results are averaged over 1000 samples of $\theta$.}
  \label{fig:static}
\end{figure}

Following the standard finite-size scaling ansatz, in the critical region near the critical point with a fixed non-Hermitian parameter $h$, the three observables satisfy the scaling forms~\cite{Zhai2022a,Bu2022,Sun2024,Sahoo2025}
\begin{equation}
\label{staticXi}
\xi(\varepsilon, L) = L f_1\left(\varepsilon L^{1/\nu}\right),
\end{equation}
\begin{equation}
\label{staticIPR}
{\rm IPR}(\varepsilon, L) = L^{-s/\nu} f_2\left(\varepsilon L^{1/\nu}\right),
\end{equation}
\begin{equation}
\label{staticEg}
E_g(\varepsilon, L) = L^{-z} f_3\left(\varepsilon L^{1/\nu}\right),
\end{equation}
where $\varepsilon = V - V_c$, and $f_1$, $f_2$, and $f_3$ are universal scaling functions for the $\xi$, the IPR, and the $E_g$, respectively.
The critical exponents $\nu$, $s$, and $z$ are to be determined by collapsing the numerical data for different system sizes onto universal curves.

To determine the critical exponents, we numerically compute the three observables for various system sizes $L$ near the critical point at a fixed non-Hermitian parameter $h$.
The results are presented in Fig.~\ref{fig:static}.
Figure~\ref{fig:static}(a1) shows the $\xi$ as a function of $\varepsilon$ for different $L$.
Following the finite-size scaling ansatz in Eq.~\eqref{staticXi}, we rescale the data by plotting $\xi/L$ against $\varepsilon L^{1/\nu}$.
As shown in Fig.~\ref{fig:static}(a2), the best collapse of the rescaled curves is achieved at $\nu = 1.00(2)$, which thereby determines the correlation length exponent $\nu$.

Figure~\ref{fig:static}(b1) presents the $\rm IPR$ as a function of $\varepsilon$ for different system sizes $L$.
Taking the determined exponent $\nu = 1.00(2)$ as input, we rescale the data according to Eq.~\eqref{staticIPR} by plotting ${\rm IPR}L^{s/\nu}$ against $\varepsilon L^{1/\nu}$.
As shown in Fig.~\ref{fig:static}(b2), the best collapse of the rescaled curves is achieved at $s = 0.7967(3)$, which determines the critical exponent for the IPR.
Figure~\ref{fig:static}(c1) displays the $E_g$ versus $\varepsilon$ for various $L$.
With the exponent $\nu = 1.00(2)$ fixed, we rescale the data following Eq.~\eqref{staticEg} by plotting $E_g L^{z}$ as a function of $\varepsilon L^{1/\nu}$.
The optimal collapse, shown in Fig.~\ref{fig:static}(c2), is obtained at $z = 1.999(2)$, which gives the dynamical exponent $z$.

\begin{figure}[tbp]
\centering
  \includegraphics[width=2.5 in,clip]{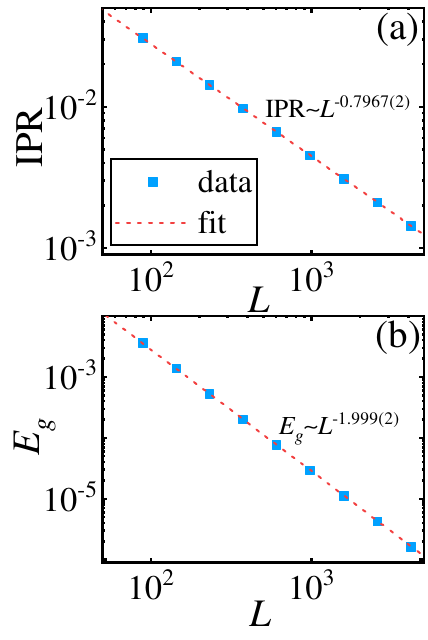}
  \vskip-3mm
  \caption{Finite-size scaling of the IPR and $E_g$ at the critical point $\varepsilon = 0$ for ($h_c = 0.693$, $V_c = 1$).
  (a) IPR versus $L$ on a log-log scale, with a power-law fit yielding ${\rm IPR} \propto L^{-0.7965(2)}$.
  (b) $E_g$ versus $L$ on a log-log scale, with a power-law fit yielding $E_g \propto L^{-1.999(2)}$.
  Results are averaged over 1000 samples of $\theta$.}
  \label{fig:EgIPRL}
\end{figure}

To further validate the extracted critical exponents, we examine the finite-size scaling of the energy gap $E_g$ and the IPR directly at the critical point $\varepsilon = 0$.
Figure~\ref{fig:EgIPRL}(a) shows $E_g$ as a function of $L$ on a double-logarithmic scale.
According to the scaling theory, $E_g \propto L^{-z}$ at the critical point.
A power-law fitting yields $E_g \propto L^{-1.999(2)}$, which is in excellent agreement with $z = 1.999(2)$ obtained from the scaling collapse.
Figure~\ref{fig:EgIPRL}(b) presents the corresponding results for the IPR, where the theory predicts ${\rm IPR} \propto L^{-s/\nu}$.
The power-law fitting gives ${\rm IPR} \propto L^{-0.7966(2)}$, which is fully consistent with $s/\nu = 0.7965(2)$ from our previous analysis.
These independent checks confirm the reliability of the critical exponents extracted from the finite-size scaling analysis.

\begin{table}[tbp]
\centering
\caption{Critical exponents for the Hermitian AA model, the nonreciprocal AA model, and the gain-loss type non-Hermitian AA model.}
\begin{tabular}{ccccc}
\hline
Model & $\nu$ & $z$ & $s$ & $r$ \\
\hline
Hermitian AA & $1$ & $2.37$ & $0.333$ & 3.37\\
Nonreciprocal AA & $1$ & $2$ & $0.1197$& 3\\
Gain-loss AA & $1.00(2)$ & $1.999(2)$ & $0.7965(2)$& 2.999\\
\hline
\end{tabular}
\label{tab:compare}
\end{table}

We now compare the extracted critical exponents with those of the nonreciprocal hopping AA model and the Hermitian AA model, as summarized in Table~\ref{tab:compare}.
For the gain-loss type non-Hermitian AA model, we obtain $\nu = 1.00(2)$ and $z = 1.999(2)$, which are almost identical to those of the nonreciprocal AA model ($\nu = 1$, $z = 2$)~\cite{Zhai2022a}, while the Hermitian AA model has $z = 2.37$~\cite{Sinha2019}.
However, the exponent $s$ exhibits a significant difference: our result $s = 0.7965(2)$ is markedly larger than both the Hermitian AA value $s = 0.333$ and the nonreciprocal AA value $s = 0.1197$~\cite{Bu2022}.
This indicates that, while the correlation length exponent $\nu$ and the dynamical exponent $z$ are nearly identical between the gain-loss and nonreciprocal models, the multifractal property of the wave functions at the critical point, as characterized by the IPR exponent $s$, is highly sensitive to the specific form of the non-Hermitian mechanism.
\section{\label{dynamics}the driven dynamics with initial extended states}
In this section, we investigate the driven dynamics of the gain-loss type non-Hermitian AA model under PBC with the initial state prepared as an extended state.
\begin{figure}[tbp]
\centering
  \includegraphics[width=2.5 in,clip]{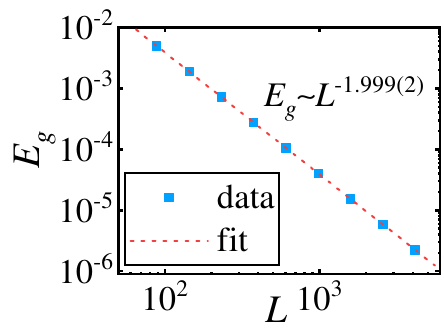}
  \vskip-3mm
  \caption{Energy gap $\Delta E$ as a function of $L$ for the extended and critical states.
  For the extended state, we use the ground state of $V = V_c - 0.5$ as an example.
  The results are averaged for $1000$ samples of $\phi$. The log-log coordinate is used.}
  \label{fig:ExtendEgL}
\end{figure}

Since the extended phase is gapless in the thermodynamic limit, we first examine the scaling of the energy gap in this regime.
As shown in Fig.~\ref{fig:ExtendEgL}, a power-law fit for the extended state yields $E_g \propto L^{-z'}$ with $z' = 1.999(2)$.
The system is then linearly driven across the critical point by varying the quasiperiodic potential strength as
\begin{equation}\label{driver}
  \varepsilon(t) = \varepsilon_0 + Rt,
\end{equation}
where $R$ is the driving rate. We focus on the dynamics of the IPR as the primary observable to characterize the nonequilibrium evolution.

\subsection{FTS for the gapless initial extended states}

We begin by summarizing the key concepts of the KZM and FTS that underpin our analysis of the driven dynamics. In the standard KZM, a system driven across a critical point at a finite rate $R$ cannot follow the instantaneous ground state arbitrarily close to the transition. The competition between the relaxation time, which diverges at the critical point, and the inverse driving rate defines a freeze-out scale $|\varepsilon| \sim R^{1/r\nu}$, where $r = z + 1/\nu$~\cite{Dziarmaga2010,Polkovnikov2011,Sinha2019}. Beyond this point, the dynamics enters an impulse regime and the system falls out of equilibrium, generating defects whose density is controlled by the driving rate.

The FTS theory~\cite{Gong2010,Huang2014} generalizes this picture by treating $R$ as a scaling variable on the same footing as the system size $L$ and the detuning $\varepsilon$. This framework provides a complete description of the entire driven evolution, unifying both the slow-driving (adiabatic) and fast-driving (impulse) limits within a single scaling form.

A crucial point for our study is that the standard KZM assumes the initial state to be gapped, so that adiabatic evolution is possible at early times. For a gapless initial state, such as an extended state in the present model, this assumption no longer holds.
However, a generalized criterion has been established~\cite{Zeng2025,Wang2026}: FTS remains applicable provided that $z' < r$, where $z'$ is the exponent governing the gap closure in the initial phase.
When this condition is met, the critical fluctuations dominate over the initial-state excitations, and the universal scaling behavior is determined by the critical point itself.

In our case, the extended phase of the gain-loss AA model has $z' = 1.999(2)$.
With the static exponents $\nu = 1.00(2)$ and $z = 1.999(2)$ from Sec.~\ref{static}, we obtain $r \approx 2.999$, which is larger than $z'$.
The FTS criterion is therefore satisfied.
In the Hermitian AA model, the FTS scaling form of the IPR has been established for driven dynamics starting from extended initial states, as derived in Ref.~\cite{Wang2026}.
Here we adopt the same scaling ansatz and generalize it to the non-Hermitian case.
The validity of this generalization will be verified by our numerical simulations in the following.

The IPR is expected to satisfy the following FTS scaling form~\cite{Wang2026}
\begin{eqnarray}
\label{Eq:ScalingIPR}
{\rm{IPR}} =\left\{ \begin{array}{l}
   L^{-\frac{s}{\nu}}g_1(RL^r,\varepsilon R^{-\frac{1}{r\nu}}), \quad \text{for small } R,  \\
  L^{-1}R^{\frac{s-\nu}{r\nu}}g_2(RL^r,\varepsilon R^{-\frac{1}{r\nu}}), \quad \text{for large } R,  \\
\end{array} \right.
\end{eqnarray}
where $g_1$ and $g_2$ are universal scaling functions that tend to constants at $\varepsilon = 0$.
The crossover between the small-$R$ and large-$R$ regimes is controlled by the scaling variable $RL^r$.
For small $R$, the driving is sufficiently slow so that the system can closely follow the instantaneous ground state throughout the evolution.
In this limit, the dynamics approaches the equilibrium regime, and the IPR in the critical region recovers the static finite-size scaling form ${\rm{IPR}} \propto L^{-s/\nu}$, consistent with Eq.~\eqref{staticIPR}.
For large $R$, the driving is too fast for the system to adjust, and the evolution retains significant memory of the initial extended state.
Consequently, the IPR exhibits a hybrid scaling behavior ${\rm{IPR}} \propto L^{-1}R^{(s-\nu)/(r\nu)}$, where the $L^{-1}$ factor reflects the extended nature of the initial state and the $R$-dependent power law arises from the nonadiabatic driving.
The two limiting behaviors are unified by Eq.~\eqref{Eq:ScalingIPR}.

\begin{figure}[tbp]
\centering
  \includegraphics[width=2.5 in,clip]{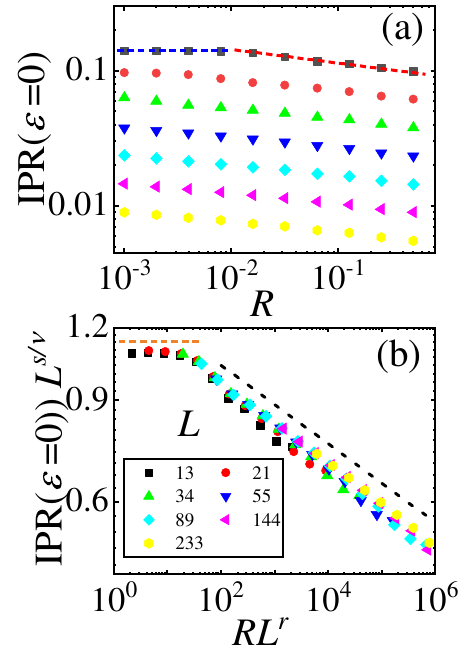}
  \vskip-3mm
  \caption{(a) IPR at the critical point $\varepsilon = 0$ as a function of the driving rate $R$ for different system sizes $L$.
  (b) Scaling collapse of the data according to the FTS form. The black dashed line has a slope of $-0.0685(3)$, consistent with the theoretical prediction $(s-\nu)/(r\nu)=-0.0679$.
  The log-log coordinate is used.}
  \label{fig:IPRvc}
\end{figure}

\subsection{numerical results}

To verify the FTS scaling form in Eq.~\eqref{Eq:ScalingIPR}, we numerically solve the time-dependent Schr\"odinger equation for the gain-loss type non-Hermitian AA model.
The system is initialized in the ground state deep in the extended phase.
The time evolution is performed using a first-order finite-difference method with a sufficiently small time step $\delta t = 1\times10^{-4}$ to ensure numerical convergence.
The IPR is calculated from the instantaneous wave function during the driving process.

To verify the scaling form in Eq.~\eqref{Eq:ScalingIPR}, we first compute the IPR at the critical point $\varepsilon = 0$ as a function of the driving rate $R$ for different system sizes $L$.
As shown in Fig.~\ref{fig:IPRvc}(a), in the log-log scale, the curves for different $L$ form a set of nearly parallel lines but with distinct slopes in the small-$R$ and large-$R$ regimes, especially for smaller system sizes (e.g., the blue and red curves in the figure).

Furthermore, as shown in Fig.~\ref{fig:IPRvc}(b), after rescaling the data according to the scaling form, the curves for different system sizes collapse onto a single universal curve, confirming the validity of the scaling equation.
It is observed that the rescaled curves of ${\rm IPR}(\varepsilon=0) L^{s/\nu}$ versus $RL^r$ separate into two distinct regimes, with the crossover occurring around $RL^r \sim 19.21$.
For $RL^r \lesssim 19.21$, the rescaled quantity ${\rm IPR}(\varepsilon=0) L^{s/\nu}$ remains approximately constant, i.e., ${\rm IPR}(\varepsilon=0) L^{s/\nu} \propto (RL^r)^0$, which implies ${\rm IPR}(\varepsilon=0) \propto L^{-s/\nu}$, thereby confirming the scaling behavior in the small-$R$ regime predicted by Eq.~\eqref{Eq:ScalingIPR}.
In the large-$R$ regime, we find that the slope of ${\rm IPR}(\varepsilon=0) L^{s/\nu}$ versus $RL^r$ is $-0.0685(3)$, which is in excellent agreement with the theoretical prediction $(s-\nu)/(r\nu) \approx -0.0679$.
This implies ${\rm IPR}(\varepsilon=0) L^{s/\nu} \propto (RL^r)^{(s-\nu)/(r\nu)}$, or equivalently ${\rm IPR}(\varepsilon=0) \propto R^{(s-\nu)/(r\nu)} L^{-1}$, thereby confirming the scaling behavior in the large-$R$ regime predicted by Eq.~\eqref{Eq:ScalingIPR}.

\begin{figure}[tbp]
\centering
  \includegraphics[width=2.6 in,clip]{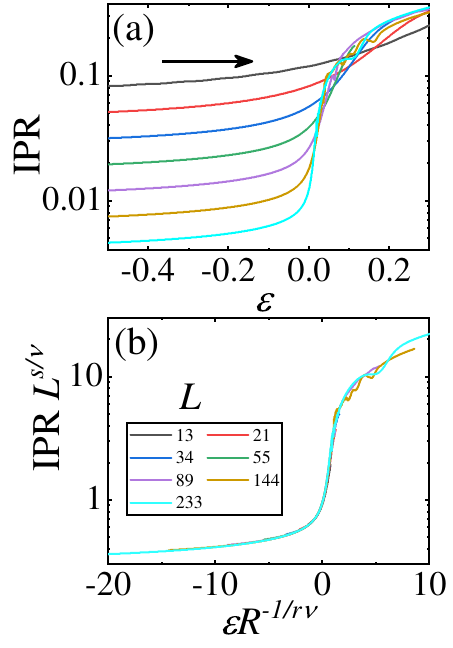}
  \vskip-3mm
  \caption{(a) IPR as a function of $\varepsilon$ for different system sizes $L$ at a fixed $RL^r$. (b) Scaling collapse of the data according to Eq.~\eqref{Eq:UnifiedScaling}.Here the scaled driving rate is fixed at $RL^r = 126.49$, results are averaged over 10 to 100 samples of $\theta$, and the $y$-axis is on a logarithmic scale.}
  \label{fig:fixRL}
\end{figure}

Subsequently, we fix the scaled driving rate $RL^r$ to a constant. In this case, the large-$R$ scaling form in Eq.~\eqref{Eq:ScalingIPR} can be rewritten as
\begin{equation}
{\rm{IPR}}(\varepsilon, R, L) = (RL^r)^{\frac{s-\nu}{r\nu}} L^{-s/\nu} g_2\left(\varepsilon R^{-1/(r\nu)}\right),
\end{equation}
where the factor $(RL^r)^{(s-\nu)/(r\nu)}$ is a constant and can be absorbed into the scaling function.
Therefore, the IPR follows the unified scaling form
\begin{equation}
{\rm{IPR}}(\varepsilon, R, L) = L^{-s/\nu} g_3\left(\varepsilon R^{-1/(r\nu)}\right),
\label{Eq:UnifiedScaling}
\end{equation}
which holds for both the small-$R$ and large-$R$ regimes.
Here $g_3$ is a universal scaling function.
This unified form indicates that when $RL^r$ is held fixed, the IPR scaling is governed solely by the static critical exponent $s/\nu$, with all dependence on the driving rate $R$ entering through the combination $\varepsilon R^{-1/(r\nu)}$.
To verify this prediction, we numerically compute the IPR for various system sizes $L$ at a fixed value of $RL^r$, and the results are presented in Fig.~\ref{fig:fixRL}.
The excellent collapse of the rescaled data confirms the validity of Eq.~\eqref{Eq:UnifiedScaling}.

\begin{figure}[tbp]
\centering
  \includegraphics[width=2.7 in,clip]{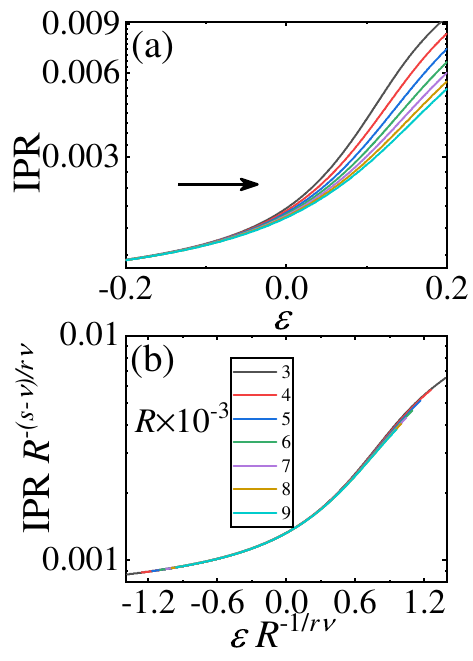}
  \vskip-3mm
  \caption{(a) IPR as a function of $\varepsilon$ for different driving rates $R$ at a fixed large system size $L = 987$. (b) Scaling collapse of the data according to Eq.~\eqref{Eq:LargeL}. Results are averaged over 10 to 100 samples of $\theta$, and the $y$-axis is on a logarithmic scale.}
  \label{fig:fixL}
\end{figure}

Finally, we examine the large-$L$ limit, where $L$ is sufficiently large that the scaling variable $RL^r \to \infty$.
In this limit, the $L^{-1}$ factor in the large-$R$ scaling form in Eq.~\eqref{Eq:ScalingIPR} disappears, and the IPR scaling reduces to
\begin{equation}
{\rm{IPR}}(\varepsilon, R) = R^{\frac{s-\nu}{r\nu}} g_4\left(\varepsilon R^{-1/(r\nu)}\right),
\label{Eq:LargeL}
\end{equation}
where $g_4$ is a universal scaling function.
To verify this prediction, we take a fixed large system size $L = 987$ and compute the IPR as a function of $\varepsilon$ for various driving rates $R$.
The results are presented in Fig.~\ref{fig:fixL}, where the rescaled curves according to Eq.~\eqref{Eq:LargeL} collapse onto a single universal curve, confirming the validity of the scaling form in the large-$L$ limit.

\section{\label{sum}conclusion}
In summary, we have systematically investigated the static critical properties and driven dynamics of the gain-loss type non-Hermitian AA model.
Through finite-size scaling analyses of the localization length $\xi$, the IPR, and the energy gap $E_g$, we have extracted the critical exponents $\nu = 1.00(2)$, $s = 0.7965(2)$, and $z = 1.999(2)$.
These exponents are distinct from those of the nonreciprocal hopping AA model, particularly the IPR exponent $s$, demonstrating that different non-Hermitian mechanisms lead to fundamentally different multifractal properties at the critical point and thus belong to distinct universality classes.

For the driven dynamics starting from a gapless extended initial state, we have verified that the FTS framework remains applicable provided that the criterion $z' < r$ is satisfied.
Using the static exponents, we have confirmed that $z' = 1.999(2) < r \approx 2.999$, and the predicted scaling forms for the IPR have been numerically validated across a wide range of system sizes and driving rates.
Our results demonstrate that the unified FTS scaling form, originally established in Hermitian systems, can be successfully generalized to the gain-loss type non-Hermitian AA model.
This extends the applicability of the FTS framework beyond both Hermitian systems and the previously studied nonreciprocal non-Hermitian models.

Our findings highlight the crucial role of the initial state in shaping the nonequilibrium dynamics and establish the gain-loss type non-Hermitian AA model as a valuable platform for exploring the interplay between non-Hermiticity, criticality, and driven dynamics.
Given the experimental accessibility of non-Hermitian quasiperiodic systems in photonic and cold-atom platforms, our predictions are expected to stimulate further experimental investigations~\cite{Lin2026,XiaoL2025,Zhang2026PRX}.
\section*{Acknowledgments}
This work is supported by the National Natural Science Foundation of China (Grant Nos. 12274184 and 12404105), the Qing Lan Project, the Natural Science Foundation of Jiangsu Province (No. BK20233001), Jiangsu Key Laboratory of Frontier Material Physics and Devices (KJS2351) and the Natural Science Foundation of the Jiangsu Higher Education Institutions of China (Grant No. 24KJB140008).


\begin{thebibliography}{120}
\bibitem{Anderson1958}P. W. Anderson, Absence of diffusion in certain random lattices, Phys. Rev. {\bf109}, 1492 (1958).
\bibitem{Thouless1974} D. J. Thouless, Electrons in disordered systems and the theory of localization, Phys. Rep. {\bf 13}, 93 (1974).
\bibitem{Abrahams1979}E. Abrahams, P.W. Anderson, D.C. Licciardello, and T.V. Ramakrishnan, Scaling Theory of Localization: Absence of Quantum Diffusion in Two Dimensions, Phys. Rev. Lett. {\bf 42}, 673-676 (1979).
\bibitem{Lee1985}P. A. Lee and T. V. Ramakrishnan, Disordered electronic systems, Rev. Mod. Phys. {\bf57}, 287 (1985).
\bibitem{Kramer1993}B. Kramer, A. MacKinnon, Localization: theory and experiment, Rep. Prog. Phys. {\bf56}, 1469 (1993).
\bibitem{Schwartz2007}T. Schwartz, G. Bartal, S. Fishman, and M. Segev, Transport and Anderson localization in disordered two-dimensional photonic lattices, Nature {\bf446}, 7131 (2007).
\bibitem{Evers2008}F. Evers, A.D. Mirlin, Anderson transitions, Rev. Mod. Phys., {\bf80}, 1355-1417 (2008).
\bibitem{Segev2013}M. Segev, Y. Silberberg, and D. N. Christodoulides, Anderson localization of light, Nature Photonics {\bf7}, 197 (2013).
\bibitem{Wiersma2013}D. S. Wiersma, Disordered photonics, Nature Photonics {\bf7}, 188 (2013).
\bibitem{Ekuma2015} C. E. Ekuma, S.-X. Yang, H. Terletska, K.-M. Tam, N. S. Vidhyadhiraja, J. Moreno, and M. Jarrell, Metal-insulator transition in a weakly interacting disordered electron system, Phys. Rev. B {\bf 92}, 201114(R) (2015).
\bibitem{Nguyen2022} T. H. Y. Nguyen, D. A. Le, and A. T. Hoang, Anderson localization in the Anderson-Hubbard model with site-dependent interactions, New J. Phys. {\bf 24}, 053054 (2022).
\bibitem{Alexey2023}A. Yamilov, S. E. Skipetrov, T. W. Hughes, M. Minkov, Z. Yu, and H. Cao, Anderson localization of electromagnetic waves in three dimensions, Nat. Phys. {\bf19}, 1308 (2023).
\bibitem{Qi2026} Z. Qi, Y. Zhang, M. Qin, H. Weng, and K. Jiang, Anderson localization: A density matrix approach, Phys. Rev. X {\bf 16}, 011043 (2026).



\bibitem{Aubry1980}S. Aubry and G. Andr\'{e}, Analyticity breaking and anderson localization in incommensurate lattices, Ann. Israel Phys. Soc. {\bf3}, 133 (1980).
\bibitem{Biddle070601}J. Biddle and S. Das Sarma, Predicted mobility edges in one-dimensional incommensurate optical lattices: An exactly solvable model of anderson localization, Phys. Rev. Lett. {\bf104}, 070601 (2010).
\bibitem{HepengYao2019}H. Yao, A. Khoudli, L. Bresque, and L. Sanchez-Palencia, Critical behavior and fractality in shallow one-dimensional quasiperiodic potentials, Phys. Rev. Lett. {\bf123}, 070405 (2019).
\bibitem{Schirmann2024}J. Schirmann, S. Franca, F. Flicker, and A. G. Grushin, Physical Properties of an Aperiodic Monotile with Graphene-like Features, Chirality, and Zero Modes, Phys. Rev. Lett. {\bf132}, 086402 (2024).
\bibitem{Ribeiro2024}M. Gon\c{c}alves, B. Amorim, E. V. Castro, and P. Ribeiro, Critical Phase Dualities in 1D Exactly Solvable Quasiperiodic Models, Phys. Rev. Lett. {\bf131}, 186303 (2024).
\bibitem{Jan2023}J. \v{S}untajs, T. c. v. Prosen, and L. Vidmar, Localization challenges quantum chaos in the finite two-dimensional Anderson model, Phys. Rev. B {\bf107}, 064205 (2023).
\bibitem{Agrawalprl2020}U. Agrawal, S. Gopalakrishnan, and R. Vasseur, Quantum Criticality in the 2D Quasiperiodic Potts Model, Phys. Rev. Lett. {\bf125}, 265702 (2020).
\bibitem{Goblot2020}V. Goblot, A. \v{S}trkalj, N. Pernet, J. L. Lado, C. Dorow, A. Lema\^{\i}tre, L. L. Gratiet, A. Harouri, I. Sagnes, S. Ravets, A. Amo, J. Bloch, and O. Zilberberg, Emergence of criticality through a cascade of delocalization transitions in quasiperiodic chains, Nat. Phys. {\bf16}, 832 (2020).
\bibitem{Roy2022}S. Roy, S. Chattopadhyay, T. Mishra, and S. Basu, Critical analysis of the reentrant localization transition in a one-dimensional dimerized quasiperiodic lattice, Phys. Rev. B {\bf105}, 214203 (2022).
\bibitem{Cheng2023} S. Cheng, R. Asgari, and G. Xianlong, From topological phase to transverse Anderson localization in a two-dimensional quasiperiodic system, Phys. Rev. B {\bf 108}, 024204 (2023).


\bibitem{DasSarma1988} S. Das Sarma, S. He, and X. C. Xie, Mobility edge in a model one-dimensional potential, Phys. Rev. Lett. {\bf 61}, 2144 (1988).
\bibitem{Biddle2009} J. Biddle, B. Wang, D. J. Priour, Jr., and S. Das Sarma, Localization in one-dimensional incommensurate lattices beyond the Aubry-Andr\'e model, Phys. Rev. A {\bf 80}, 021603(R) (2009).
\bibitem{Biddle2011} J. Biddle, D. J. Priour, B. Wang, and S. Das Sarma, Localization in one-dimensional lattices with non-nearest-neighbor hopping: Generalized Anderson and Aubry-Andr\'e models, Phys. Rev. B {\bf 83}, 075105 (2011).
\bibitem{Ganeshan2015} S. Ganeshan, J. H. Pixley, and S. Das Sarma, Nearest neighbor tight binding models with an exact mobility edge in one dimension, Phys. Rev. Lett. {\bf 114}, 146601 (2015).
\bibitem{Liu2015} F. Liu, S. Ghosh, and Y. D. Chong, Localization and adiabatic pumping in a generalized Aubry-Andr\'e-Harper model, Phys. Rev. B {\bf 91}, 014108 (2015).
\bibitem{Wang2020} Y. Wang, X. Xia, L. Zhang, H. Yao, S. Chen, J. You, Q. Zhou, and X.-J. Liu, One-dimensional quasiperiodic mosaic lattice with exact mobility edges, Phys. Rev. Lett. {\bf 125}, 196604 (2020).
\bibitem{Zhou2023} X.-C. Zhou, Y.-J. Wang, T.-F. J. Poon, Q. Zhou, and X.-J. Liu, Exact new mobility edges between critical and localized states, Phys. Rev. Lett. {\bf 131}, 176401 (2023).
\bibitem{Hetenyi2025} B. Het\'enyi and I. Balogh, Numerical study of the localization transition of Aubry-Andr\'e type models, Phys. Rev. B {\bf 112}, 144203 (2025).
\bibitem{Wangyn2026} Y. N. Wang, W. L. You, Z. Xu, and G. Sun, Generalized Aubry-Andr\'e-Harper model with power-law quasiperiodic potentials, Phys. Rev. B {\bf 113}, 224204 (2026).
\bibitem{Bu2022} X. Bu, L.-J. Zhai, and S. Yin, Quantum criticality in the disordered Aubry-Andr\'e model, Phys. Rev. B {\bf 106}, 214208 (2022).

\bibitem{DuL2025} L. Du, Y. Liu, M. Shi, S. Wang, Y. He, X. Zhou, X. Yang, L. Tao, K. Song, Z. Li, and X. Zhao, Experimental observation of interconversion of non-Hermitian skin effect and Anderson localization in a nonreciprocal topological circuit based on Aubry-Andr\'e model, Phys. Rev. B {\bf 112}, 115417 (2025).
\bibitem{Roati2008} G. Roati, C. D'Errico, L. Fallani, M. Fattori, C. Fort, M. Zaccanti, G. Modugno, M. Modugno, and M. Inguscio, Anderson localization of a non-interacting Bose-Einstein condensate, Nature (London) {\bf 453}, 895 (2008).
\bibitem{Billy2008} J. Billy, V. Josse, Z. Zuo, A. Bernard, B. Hambrecht, P. Lugan, D. Cl\'ement, L. Sanchez-Palencia, P. Bouyer, and A. Aspect, Direct observation of Anderson localization of matter waves in a controlled disorder, Nature (London) {\bf 453}, 891 (2008).
\bibitem{Lahini2009} Y. Lahini, R. Pugatch, F. Pozzi, M. Sorel, R. Morandotti, N. Davidson, and Y. Silberberg, Observation of a localization transition in quasiperiodic photonic lattices, Phys. Rev. Lett. {\bf 103}, 013901 (2009).
\bibitem{Schreiber2015} M. Schreiber, S. S. Hodgman, P. Bordia, H. P. L\"uschen, M. H. Fischer, R. Vosk, E. Altman, U. Schneider, and I. Bloch, Observation of many-body localization of interacting fermions in a quasirandom optical lattice, Science {\bf 349}, 842 (2015).
\bibitem{Luschen2018} H. P. L\"uschen, S. Scherg, T. Kohlert, M. Schreiber, P. Bordia, X. Li, S. Das Sarma, and I. Bloch, Single-particle mobility edge in a one-dimensional quasiperiodic optical lattice, Phys. Rev. Lett. {\bf 120}, 160404 (2018).
\bibitem{Kohlert2019} T. Kohlert, S. Scherg, X. Li, H. P. L\"uschen, S. Das Sarma, I. Bloch, and M. Aidelsburger, Observation of many-body localization in a one-dimensional system with a single-particle mobility edge, Phys. Rev. Lett. {\bf 122}, 170403 (2019).
\bibitem{An2021} F. A. An, K. Padavi\'c, E. J. Meier, S. Hegde, S. Ganeshan, J. H. Pixley, S. Vishveshwara, and B. Gadway, Interactions and mobility edges: Observing the generalized Aubry-Andr\'e model, Phys. Rev. Lett. {\bf 126}, 040603 (2021).
\bibitem{Rispoli2019} M. Rispoli, A. Lukin, R. Schittko, S. Kim, M. E. Tai, J. L\'eonard, and M. Greiner, Quantum critical behaviour at the many-body localization transition, Nature (London) {\bf 573}, 385 (2019).
\bibitem{Singh2015} K. Singh, K. Saha, S. A. Parameswaran, and D. M. Weld, Fibonacci optical lattices for tunable quantum quasicrystals, Phys. Rev. A {\bf 92}, 063426 (2015).
\bibitem{Xiao2021} L. Xiao, D.-K. Qu, K.-K. Wang, H.-W. Li, J.-Y. Dai, B. D\'ora, M. Heyl, R. Moessner, W. Yi, and P. Xue, Non-Hermitian Kibble-Zurek mechanism with tunable complexity in single-photon interferometry, PRX Quantum {\bf 2}, 020313 (2021).
\bibitem{Lin2022} Q. Lin, T.-Y. Li, L. Xiao, K.-K. Wang, W. Yi, and P. Xue, Observation of non-Hermitian topological Anderson insulator in quantum dynamics, Nat. Commun. {\bf 13}, 3229 (2022).



\bibitem{Hatano1996} N. Hatano and D. R. Nelson, Localization transitions in non-Hermitian quantum mechanics, Phys. Rev. Lett. {\bf 77}, 570 (1996).
\bibitem{Hatano1998} N. Hatano and D. R. Nelson, Non-Hermitian delocalization and eigenfunctions, Phys. Rev. B {\bf 58}, 8384 (1998).
\bibitem{Jiang2019} H. Jiang, L.-J. Lang, C. Yang, S.-L. Zhu, and S. Chen, Interplay of non-Hermitian skin effects and Anderson localization in nonreciprocal quasiperiodic lattices, Phys. Rev. B {\bf 100}, 054301 (2019).
\bibitem{Zhai2021} L.-J. Zhai, G.-Y. Huang, and S. Yin, Cascade of the delocalization transition in a non-Hermitian interpolating Aubry-Andr\'e-Fibonacci chain, Phys. Rev. B {\bf 104}, 014202 (2021).
\bibitem{Sun2024} Y.-M. Sun, X.-Y. Wang, and L.-J. Zhai, Hybrid scaling properties of the localization transition in a non-Hermitian disordered Aubry-Andr\'e model, Phys. Rev. B {\bf 110}, 054202 (2024).

\bibitem{Longhi2019a} S. Longhi, Topological phase transition in non-Hermitian quasicrystals, Phys. Rev. Lett. {\bf 122}, 237601 (2019).
\bibitem{Longhi2019b} S. Longhi, Metal-insulator phase transition in a non-Hermitian Aubry-Andr\'e-Harper model, Phys. Rev. B {\bf 100}, 125157 (2019).
\bibitem{Zeng2017} Q.-B. Zeng, S. Chen, and R. L\"u, Anderson localization in the non-Hermitian Aubry-Andr\'e-Harper model with physical gain and loss, Phys. Rev. A {\bf 95}, 062118 (2017).
\bibitem{Xu2021} Z. Xu and S. Chen, Dynamical evolution in a one-dimensional incommensurate lattice with PT symmetry, Phys. Rev. A {\bf 103}, 043325 (2021).
\bibitem{Xing2025} Z.-Y. Xing, S. Chen, and H. Hu, Universal spreading dynamics in quasiperiodic non-Hermitian systems, Phys. Rev. B {\bf 111}, L180203 (2025).
\bibitem{Cai2021} X. Cai, Anderson localization and topological phase transitions in non-Hermitian Aubry-Andr\'e-Harper models with p-wave pairing, Phys. Rev. B {\bf 103}, 214202 (2021).
\bibitem{Ghatak2024} A. Ghatak, D. H. Kaltsas, M. Kulkarni, and K. G. Makris, Diffraction and pseudospectra in non-Hermitian quasiperiodic lattices, Phys. Rev. E {\bf 110}, 064228 (2024).
\bibitem{RoyS2022} S. Roy, S. K. Maiti, L. M. P\'{e}rez, J. H. Ojeda Silva, and D. Laroze, Localization Properties of a Quasiperiodic Ladder under Physical Gain and Loss: Tuning of Critical Points, Mixed-Phase Zone and Mobility Edge, Materials {\bf 15}, 597 (2022).



\bibitem{Wang2025} Y.-P. Wang, C.-K. Chang, R. Okugawa, and C.-H. Hsu, Quasiperiodicity-induced bulk localization with self-similarity in non-Hermitian systems, Phys. Rev. B {\bf 112}, 054202 (2025).
\bibitem{Chen2025} R.-J. Chen, G.-Q. Zhang, Z. Li, and D.-W. Zhang, Mobility rings in a non-Hermitian non-Abelian quasiperiodic lattice, Phys. Rev. A {\bf 112}, 013320 (2025).
\bibitem{Zhai2020} L.-J. Zhai, S. Yin, and G.-Y. Huang, Many-body localization in a non-Hermitian quasiperiodic system, Phys. Rev. B {\bf 102}, 064206 (2020).
\bibitem{Zhang2025} Z.-Q. Zhang, H. Liu, H. Liu, H. Jiang, and X. C. Xie, Bulk-boundary correspondence in disordered non-Hermitian systems, Sci. Bull. {\bf 68}, 157 (2023).
\bibitem{Luo2021} X. Luo, T. Ohtsuki, and R. Shindou, Universality classes of the Anderson transitions driven by non-Hermitian disorder, Phys. Rev. Lett. {\bf 126}, 090402 (2021).
\bibitem{Huang2020} Y. Huang and B. I. Shklovskii, Anderson localization in non-Hermitian systems, Phys. Rev. B {\bf 101}, 014204 (2020).
\bibitem{Guo2021} C.-X. Guo, C.-H. Liu, X.-M. Zhao, Y. Liu, and S. Chen, Exact solution of non-Hermitian systems with generalized boundary conditions: Size-dependent boundary effect and fragility of the skin effect, Phys. Rev. Lett. {\bf 127}, 116801 (2021).
\bibitem{Longhi2025} S. Longhi, Erratic non-Hermitian skin localization, Phys. Rev. Lett. {\bf 134}, 196302 (2025).
\bibitem{LiB2025} B. Li, C. Chen, and Z. Wang, Universal non-Hermitian transport in disordered systems, Phys. Rev. Lett. {\bf 135}, 033802 (2025).

\bibitem{Ashida2020} Y. Ashida, Z. Gong, and M. Ueda, Non-Hermitian physics, Adv. Phys. {\bf 69}, 249 (2020).
\bibitem{Bergholtz2021} E. J. Bergholtz, J. C. Budich, and F. K. Kunst, Exceptional topology of non-Hermitian systems, Rev. Mod. Phys. {\bf 93}, 015005 (2021).
\bibitem{El-Ganainy2018} R. El-Ganainy, K. G. Makris, M. Khajavikhan, Z. H. Musslimani, S. Rotter, and D. N. Christodoulides, Non-Hermitian physics and PT symmetry, Nat. Phys. {\bf 14}, 11 (2018).
\bibitem{Yao2018} S. Yao and Z. Wang, Edge states and topological invariants of non-Hermitian systems, Phys. Rev. Lett. {\bf 121}, 086803 (2018).
\bibitem{Gong2018} Z. Gong, Y. Ashida, K. Kawabata, K. Takasan, S. Higashikawa, and M. Ueda, Topological phases of non-Hermitian systems, Phys. Rev. X {\bf 8}, 031079 (2018).
\bibitem{Kunst2018} F. K. Kunst, E. Edvardsson, J. C. Budich, and E. J. Bergholtz, Biorthogonal bulk-boundary correspondence in non-Hermitian systems, Phys. Rev. Lett. {\bf 121}, 026808 (2018).

\bibitem{Ashida2020} Y. Ashida, Z. Gong, and M. Ueda, Non-Hermitian physics, Adv. Phys. {\bf 69}, 249 (2020).
\bibitem{Bergholtz2021} E. J. Bergholtz, J. C. Budich, and F. K. Kunst, Exceptional topology of non-Hermitian systems, Rev. Mod. Phys. {\bf 93}, 015005 (2021).
\bibitem{El-Ganainy2018} R. El-Ganainy, K. G. Makris, M. Khajavikhan, Z. H. Musslimani, S. Rotter, and D. N. Christodoulides, Non-Hermitian physics and PT symmetry, Nat. Phys. {\bf 14}, 11 (2018).
\bibitem{Longhi2017} S. Longhi, Parity-time symmetry meets photonics: A new twist in non-Hermitian optics, Europhys. Lett. {\bf 120}, 64001 (2017).

\bibitem{Shen2018} H. Shen, B. Zhen, and L. Fu, Topological band theory for non-Hermitian Hamiltonians, Phys. Rev. Lett. {\bf 120}, 146402 (2018).
\bibitem{Kawabata2019} K. Kawabata, K. Shiozaki, M. Ueda, and M. Sato, Symmetry and topology in non-Hermitian physics, Phys. Rev. X {\bf 9}, 041015 (2019).
\bibitem{Xue2026} P. Xue, Essay: Topological phases and exceptional points in non-Hermitian systems, Phys. Rev. Lett. {\bf 136}, 170001 (2026).
\bibitem{Zhang2026} Q. Zhang, L. Xiong, S. Tong, and C. Qiu, Harmonic non-Hermitian skin effect, Nat. Commun. {\bf 17}, 2198 (2026).
\bibitem{Gohsrich2025} J. T. Gohsrich, A. Banerjee, and F. K. Kunst, The non-Hermitian skin effect: A perspective, Europhys. Lett. {\bf 150}, 60001 (2025).

\bibitem{Dziarmaga2010} J. Dziarmaga, Dynamics of a quantum phase transition and relaxation to a steady state, Adv. Phys. {\bf 59}, 1063 (2010).
\bibitem{Polkovnikov2011} A. Polkovnikov, K. Sengupta, A. Silva, and M. Vengalattore, Colloquium: Nonequilibrium dynamics of closed interacting quantum systems, Rev. Mod. Phys. {\bf 83}, 863 (2011).
\bibitem{Gong2010} S. Gong, F. Zhong, X. Huang, and S. Fan, Finite-time scaling via linear driving, New J. Phys. {\bf 12}, 043036 (2010).
\bibitem{Huang2014} Y. Huang, S. Yin, B. Feng, and F. Zhong, Kibble-Zurek mechanism and finite-time scaling, Phys. Rev. B {\bf 90}, 134108 (2014).
\bibitem{WangX2024} X. Wang and J. Wang, Mpemba effects in nonequilibrium open quantum systems, Phys. Rev. Res. {\bf 6}, 033330 (2024).
\bibitem{Gutierrez2023} R. Guti\'errez and R. Cuerno, Nonequilibrium criticality driven by Kardar-Parisi-Zhang fluctuations in the synchronization of oscillator lattices, Phys. Rev. Res. {\bf 5}, 023047 (2023).
\bibitem{SotoGarcia2026} J. Soto-Garcia and N. Chepiga, Quantum Kibble-Zurek mechanism: The role of boundary conditions, endpoints, and kink types, Phys. Rev. B {\bf 113}, 085430 (2026).
\bibitem{Chang2026} W.-X. Chang, S. Yin, S.-X. Zhang, and Z.-X. Li, Imaginary-time Mpemba effect in quantum many-body systems, Phys. Rev. Lett. {\bf 136}, 100403 (2026).

 \bibitem{Yang2017} C. Yang, Y. Wang, P. Wang, X. Gao, and S. Chen, Dynamical signature of localization-delocalization transition in a one-dimensional incommensurate lattice, Phys. Rev. B {\bf 95}, 184201 (2017).

\bibitem{Xu2020} Z. Xu, H. Huangfu, Y. Zhang, and S. Chen, Dynamical observation of mobility edges in one-dimensional incommensurate optical lattices, New J. Phys. {\bf 22}, 013036 (2020).

\bibitem{Morales-Molina2014} L. Morales-Molina, E. Doerner, C. Danieli, and S. Flach, Resonant extended states in driven quasiperiodic lattices: Aubry-Andr\'e localization by design, Phys. Rev. A {\bf 90}, 043630 (2014).

\bibitem{Bairey2017} E. Bairey, G. Refael, and N. H. Lindner, Driving induced many-body localization, Phys. Rev. B {\bf 96}, 020201(R) (2017).

\bibitem{Modak2021} R. Modak and D. Rakshit, Many-body dynamical phase transition in a quasiperiodic potential, Phys. Rev. B {\bf 103}, 224310 (2021).



\bibitem{Reichhardt2022} C. J. O. Reichhardt, A. del Campo, and C. Reichhardt, Kibble-Zurek mechanism for nonequilibrium phase transitions in driven systems with quenched disorder, Commun. Phys. {\bf 5}, 173 (2022).
\bibitem{Sinha2019} A. Sinha, M. M. Rams, and J. Dziarmaga, Kibble-Zurek mechanism with a single particle: Dynamics of the localization-delocalization transition in the Aubry-Andr\'e model, Phys. Rev. B {\bf 99}, 094203 (2019).
\bibitem{Tong2021} X. Tong, Y.-M. Meng, X. Jiang, C. Lee, G. D. de Moraes Neto, and G. Xianlong, Dynamics of a quantum phase transition in the Aubry-Andr\'e-Harper model with p-wave superconductivity, Phys. Rev. B {\bf 103}, 104202 (2021).


\bibitem{Bu2023} X. Bu, L.-J. Zhai, and S. Yin, Kibble-Zurek scaling in one-dimensional localization transitions, Phys. Rev. A {\bf 108}, 023312 (2023).
\bibitem{Liang2024} E.-W. Liang, L.-Z. Tang, and D.-W. Zhang, Quantum criticality and Kibble-Zurek scaling in the Aubry-Andr\'e-Stark model, Phys. Rev. B {\bf 110}, 024207 (2024).

\bibitem{Wang2026} X.-Y. Wang, W.-J. Yu, Y.-M. Sun, and L.-J. Zhai, Driven dynamics of localization phase transition in the Aubry-Andr\'e model with initial gapless extended states, Phys. Rev. Research {\bf 8}, 013093 (2026).

\bibitem{Li2024} S.-Z. Li, X.-J. Yu, and Z. Li, Emergent entanglement phase transitions in non-Hermitian Aubry-Andr\'e-Harper chains, Phys. Rev. B {\bf 109}, 024306 (2024).
\bibitem{Cheng2024} J.-Q. Cheng, S. Yin, and D.-X. Yao, Dynamical localization transition in the non-Hermitian lattice gauge theory, Commun. Phys. {\bf 7}, 58 (2024).


\bibitem{Zhai2022a} L.-J. Zhai, G.-Y. Huang, and S. Yin, Nonequilibrium dynamics of the localization-delocalization transition in the non-Hermitian Aubry-Andr\'e model, Phys. Rev. B {\bf 106}, 014204 (2022).
\bibitem{Zhai2022b} L.-J. Zhai, L.-L. Hou, Q. Gao, and H.-Y. Wang, Kibble-Zurek scaling of the dynamical localization-skin effect phase transition in a non-Hermitian quasi-periodic system under the open boundary condition, Front. Phys. {\bf 10}, 1098551 (2022).
\bibitem{Sun2025a} Y.-M. Sun, X.-Y. Wang, and L.-J. Zhai, Nonequilibrium dynamics of localization phase transition in the non-Hermitian disorder-Aubry-Andr\'e model, Phys. Rev. A {\bf 112}, 022204 (2025).

\bibitem{Bauer1990} J. Bauer, T. M. Chang, and J. L. Skinner, Correlation length and inverse-participation-ratio exponents and multifractal structure for Anderson localization, Phys. Rev. B {\bf 42}, 8121 (1990).
\bibitem{Fyodorov1992} Y. V. Fyodorov and A. D. Mirlin, Analytical derivation of the scaling law for the inverse participation ratio in quasi-one-dimensional disordered systems, Phys. Rev. Lett. {\bf 69}, 1093 (1992).

\bibitem{Wei2019} B.-B. Wei, Fidelity susceptibility in one-dimensional disordered lattice models, Phys. Rev. A {\bf 99}, 042117 (2019).

\bibitem{LiSZ2024} S.-Z. Li, E. Cheng, S.-L. Zhu, and Z. Li, Asymmetric transfer matrix analysis of Lyapunov exponents in one-dimensional nonreciprocal quasicrystals, Phys. Rev. B {\bf 110}, 134203 (2024).
\bibitem{Tong2025} X. Tong, Y. Zhang, B. Li, and X. Yang, Impact of nonreciprocal hopping on localization in non-Hermitian quasiperiodic systems, Phys. Rev. B {\bf 111}, 214202 (2025).
\bibitem{Sahoo2025} A. Sahoo, A. Saha, and D. Rakshit, Stark localization near Aubry-Andr\'e criticality, Phys. Rev. B {\bf 111}, 024205 (2025).

\bibitem{Zeng2025} Z. Zeng, Y.-K. Yu, Z.-X. Li, Z.-X. Li, and S. Yin, Finite-time scaling beyond the Kibble-Zurek prerequisite in Dirac systems, Nat. Commun. {\bf 16}, 6181 (2025).

\bibitem{Lin2026} Q. Lin, C. Cedzich, Q. Zhou, and P. Xue, Observation of metal-insulator and spectral phase transitions in Aubry-Andr\'e-Harper models, Phys. Rev. Lett. {\bf 136}, 206602 (2026).
\bibitem{XiaoL2025} L. Xiao, K. Wang, D. Qu, H. Gao, Q. Lin, Z. Bian, X. Zhan, and P. Xue, Non-Hermitian physics in photonic systems, Photonics Insights {\bf 4}, R09 (2025).
\bibitem{Zhang2026PRX} J. Zhang, Y.-J. Wang, S.-Y. Shao, B. Liu, L.-H. Zhang, Z.-Y. Zhang, X. Liu, C. Yu, Q. Li, H.-C. Chen, Y. Ma, T.-Y. Han, Q.-F. Wang, J.-D. Nan, Y.-M. Yin, D.-Y. Zhu, Q.-Q. Fang, D.-S. Ding, and B.-S. Shi, Observation of non-Hermitian topology in cold Rydberg quantum gases, PRX Quantum {\bf 7}, 033004 (2026).


\end{thebibliography}
\end{document}